\documentclass[conference]{IEEEtran}
\IEEEoverridecommandlockouts
\usepackage{cite}
\usepackage{amsmath,amssymb,amsfonts}
\usepackage{algorithmic}
\usepackage{graphicx}
\usepackage{textcomp}
\usepackage{xcolor}
\usepackage{multirow}
\usepackage{booktabs}
\usepackage{hyperref} 
\def\BibTeX{{\rm B\kern-.05em{\sc i\kern-.025em b}\kern-.08em
    T\kern-.1667em\lower.7ex\hbox{E}\kern-.125emX}}
\begin{document}

\title{HES-UNet: A U-Net for Hepatic Echinococcosis Lesion Segmentation}

\author{\IEEEauthorblockN{Jiayan Chen$^{1,2}$, Kai Li$^{3}$, Zhanjin Wang$^{4,5}$, Zhan Wang$^{4,5}$, Jianqiang Huang$^{1,2,*}$\\}
\IEEEauthorblockA{1. Department of Computer Technology and Applications, Qinghai University, Xining, China \\
2. Intelligent Computing and Application Laboratory of Qinghai Province, Qinghai University, Xining, China \\
3. Department of Computer Science  and Technology, Tsinghua University, Beijing, China \\
4. Department of General Surgery, Affiliated Hospital of Qinghai University, Xining, China \\
5. Department of Medical Engineering Integration 
and Translational Application, \\Affiliated Hospital of Qinghai University, Xining, China \\}
\thanks{$*$ Corresponding author.}
}

\maketitle
% 不要在标题中使用缩写
% 除了abstract，其他首次使用时，也不要用缩写
% 引用方程时使用（1）而不是Eq. (1) 或 equation (1)，除了在句首，可以写Equation (1) is ..

\begin{abstract}
% 100~150 words
Hepatic echinococcosis (HE) is a prevalent disease in economically underdeveloped pastoral areas, where adequate medical resources are usually lacking. Existing methods often ignore multi-scale feature fusion or focus only on feature fusion between adjacent levels, which may lead to insufficient feature fusion. To address these issues, we propose HES-UNet, an efficient and accurate model for HE lesion segmentation. This model combines convolutional layers and attention modules to capture local and global features. During downsampling, the multi-directional downsampling block (MDB) is employed to integrate high-frequency and low-frequency features, effectively extracting image details. The multi-scale aggregation block (MAB) aggregates multi-scale feature information. In contrast, the multi-scale upsampling Block (MUB) learns highly abstract features and supplies this information to the skip connection module to fuse multi-scale features. 
Due to the distinct regional characteristics of HE, there is currently no publicly available high-quality dataset for training our model. 
We collected CT slice data from 268 patients at a certain hospital to train and evaluate the model. 
The experimental results show that HES-UNet achieves state-of-the-art performance on our dataset, achieving an overall Dice Similarity Coefficient (DSC) of 89.21\%, which is 1.09\% higher than that of TransUNet. The project page is available at \url{https://chenjiayan-qhu.github.io/HES-UNet-page/}.
\end{abstract}


\section{Introduction}
\label{sec:intro}

Hepatic echinococcosis (HE) is a severe zoonotic disease caused by parasitic infection, typically occurring in remote pastoral or high-altitude regions with harsh climate conditions and limited medical resources\cite{czermak2008echinococcosis}. Computed tomography (CT) is generally the primary means of diagnosing HE\cite{filippou2007advances}; however, manual segmentation of lesions is time-consuming and reliant on subjective judgments based on physician experience, making it difficult to replicate. Early medical image segmentation methods typically relied on traditional edge detection\cite{canny1986edgedetection} and region-growing algorithms\cite{adams1994seededregiongrowing}. However, these traditional methods usually depend on hand-crafted features and are less robust to variations in lesion shape and size, making it challenging to handle the irregular shapes and blurred boundaries of HE lesions. 

In recent years, with the rapid development of deep learning, convolutional neural networks (CNNs)\cite{cnn} have shown great potential in the field of medical image segmentation. U-Net\cite{unet} is a widely used network architecture. However, due to the local receptive field inherent in convolution operations, the model struggles to effectively capture global image features. Additionally, the max-pooling operation still leads to the loss of fine details, which is critical for abdominal CT images with unclear boundaries, limiting the model’s ability to capture subtle features. Trans-UNet\cite{transunet} combines Transformer\cite{transformer} and CNN, utilizing self-attention mechanisms to capture global features, but this significantly increases the number of parameters, making the model more challenging to train. Additionally, training such a complex model requires a large amount of annotated data, which is often difficult to obtain in medical applications. Swin-UNet\cite{swinunet} proposes a U-shaped architecture using pure Transformers without relying on convolutions for feature extraction. While this design excels at modeling global features, it reduces the model's ability to learn local features, especially when dealing with fine boundaries and texture details in medical images. Furthermore, simplistic skip connections fail to effectively aggregate features across hierarchical levels.

Therefore, to address the aforementioned issues, we propose HES-UNet, an efficient and accurate segmentation model for HE that precisely delineates lesion areas to assist physicians in disease diagnosis while achieving advanced accuracy. This model is composed of four main components: a multi-scale feature integration encoder (MFSI encoder), a multi-scale global feature filtering module (MGF module), a progressive fusion decoder (PF decoder), and a deep supervision module (DS module). In the encoder section, we replace the traditional max-pooling in the U-Net with an MDB, which effectively preserves the multi-frequency features within the image. Next, we use the MAB module to aggregate global feature representations. Following this, we apply the MUB to extract multi-scale global features, which selectively enhances features at different scales. Unlike traditional skip connections between adjacent layers, MUB enables global adjustment across all feature levels. Finally, the PF decoder generates the prediction. Additionally, we introduce deep supervision in the DS module, allowing each decoder block (DB) to directly output prediction probabilities, which are then integrated into the MGF. This approach significantly enhances the model’s generalization ability and optimizes gradient propagation.

Our contributions can be summarized as follows: 1. We propose a model specifically designed for liver echinococcosis segmentation. 2. We propose three modules for enhancing segmentation features. MDB implements lossless feature downsampling, MAB facilitates the formation of global feature representations, and MUB optimizes skip connections by selectively enhancing features at different scales to provide multi-scale global features. 3. We conducted extensive experiments to demonstrate the advanced performance of our model. 
% The results showed that our model is very effective in accurately segmenting the lesion area, achieving a DSC of 90.19\%.
Our model achieved a DSC of 89.21\%, demonstrating its effectiveness in lesion segmentation.

\renewcommand{\dblfloatpagefraction}{.8}
% 添加这一行使得页面图片可以占80% % 
\begin{figure*}
\centering
\includegraphics[width=0.80\textwidth]{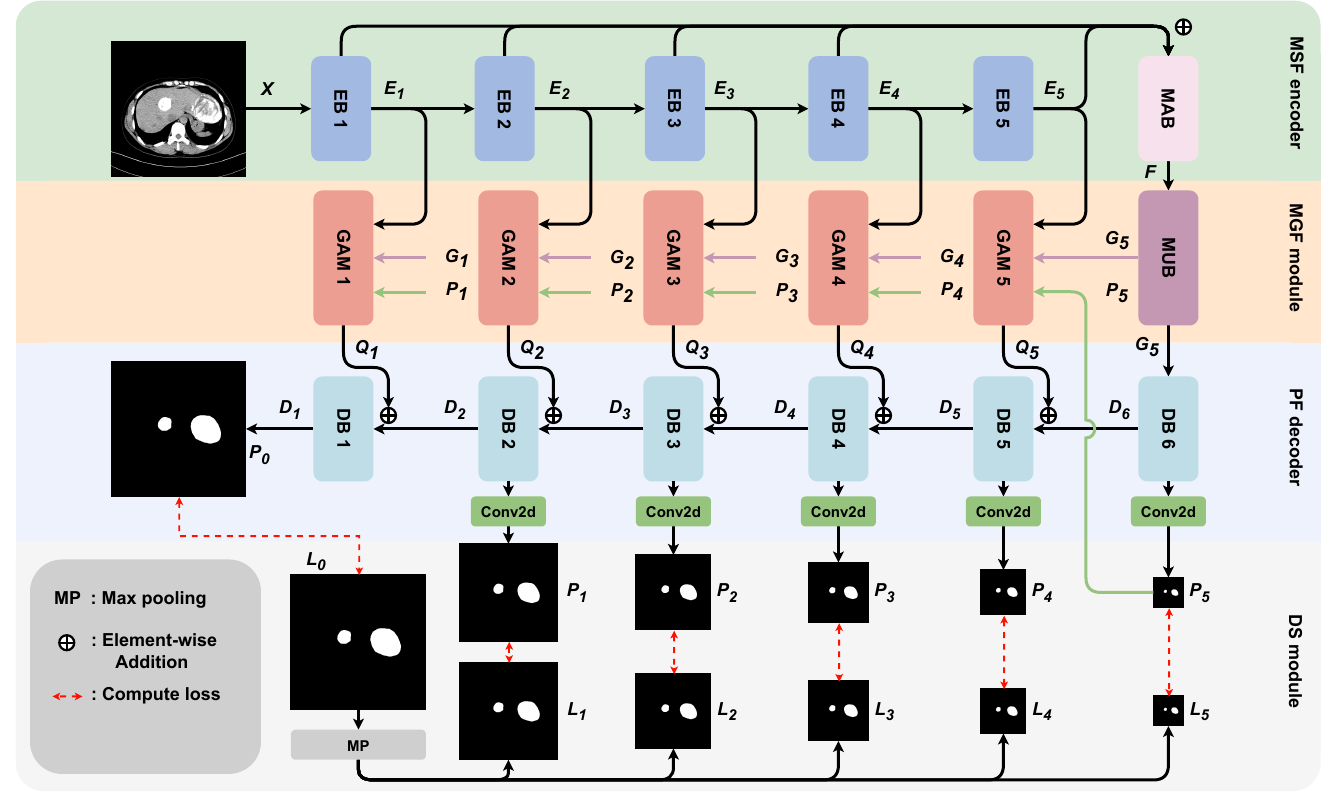}
\caption{The overall pipeline of HES-UNet. HES-UNet consists of four main components: MFSI encoder, MGF module, PF decoder, and DS module.}
\label{fig1:env}
\end{figure*}

\section{Related Work}

% \textbf{CNN-based Segmentation Modules.} 
\subsection{CNN-based Segmentation Models}

Designing a high-performance network architecture has been one of the most important research directions in the field of computer vision. Since the introduction of CNNs, they have been widely applied in medical image segmentation. Among them, U-Net\cite{unet} and its variants (such as 3D U-Net\cite{3dunet}, V-Net\cite{vnet} and U-Net++\cite{unet++}) have become the most influential models due to their symmetric encoder-decoder structure, which enables pixel-level prediction through multi-scale feature extraction and feature fusion via skip connections. However, traditional CNNs are limited by the nature of convolutional kernels, which can only extract local feature information from images and are insufficient for capturing long-range dependencies and global information\cite{ghorbani2024msa2net}. Additionally, simply using max-pooling operations during downsampling may lead to the loss of detailed information, reducing the segmentation accuracy of lesion boundaries\cite{hwd}.

\subsection{Transformer-based Segmentation Models}

Transformers were initially proposed by Vaswani et al\cite{transformer}. for sequence-to-sequence tasks, with the core idea being that the self-attention mechanism can weigh each position in the input sequence, thereby capturing long-range dependencies. The ViT\cite{vit} was the first to apply the Transformer architecture to image processing tasks, effectively learning global features in images while avoiding the drawbacks of convolution. Cao et al. proposed Swin-UNet\cite{swinunet}, which utilizes the Swin Transformer as the base architecture and includes a pair of encoder-decoder structures. The architecture employs a hierarchical self-attention mechanism to learn image features.

\subsection{Hybrid Segmentation Models}

Due to the advantages of Transformer models in capturing global features, while CNNs excel in capturing local features, researchers have begun to explore hybrid architectures that combine Transformers and CNNs to achieve more accurate medical image segmentation. Ou et al. proposed Trans-UNet\cite{transunet}, which extracts local features through CNNs and enhances global feature modeling through Transformer modules, thereby improving segmentation accuracy. Attention U-Net adds an attention mechanism to the traditional U-Net, dynamically adjusting the feature map fusion process by learning the importance of features.

\begin{figure*}[h]
\centering
\includegraphics[width=0.9\textwidth]{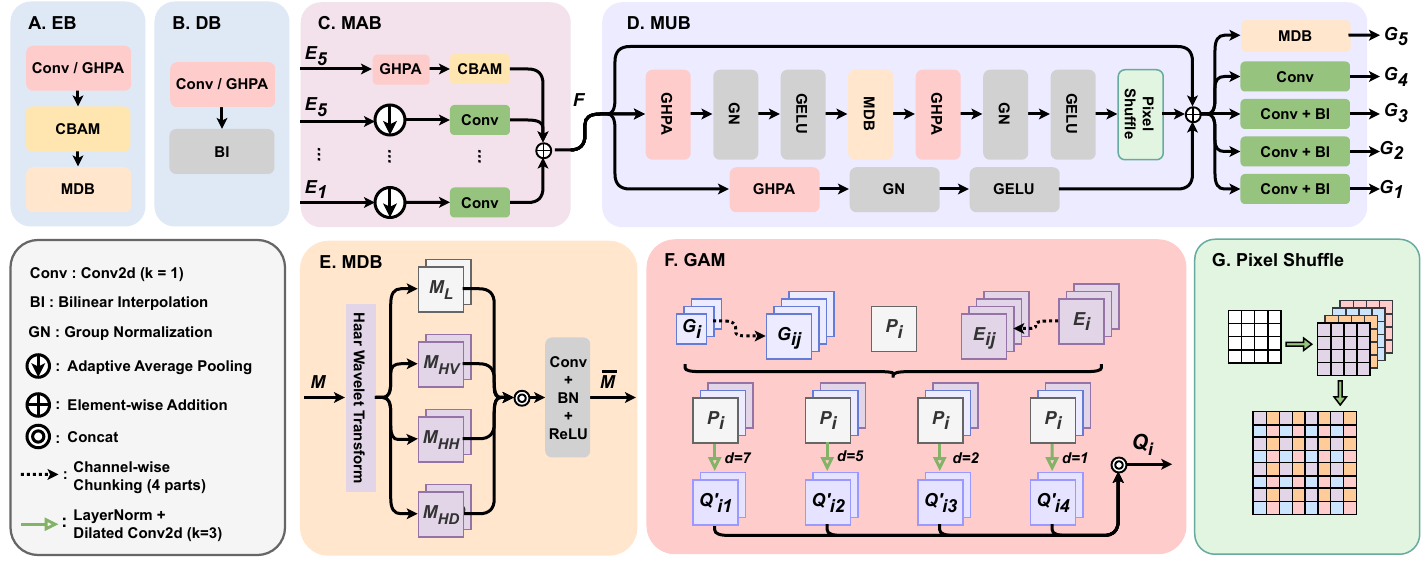}
\caption{The structure of each module in HES-UNet: (A) Encoder Block, (B) Decoder Block, (C) Multi-scale Aggregation Block, (D) Multi-scale Upsampling Block, (E) Multi-directional Downsampling Block, (F) Global Attention Module, and (G) Pixel Shuffle.}
\label{fig2:env}
\end{figure*}

\section{Methods}

HE lesion segmentation typically involves the following challenges: (1) CT images often contain noise and have low contrast; (2) HE can be classified into cystic echinococcosis (CE) and alveolar echinococcosis (AE), and the lesions of these two types differ significantly, making it challenging for the model to simultaneously learn both the common and distinct features. (3) The death of the parasite can result in calcifications in the lesion area, which appear as high-density, well-defined white spots in the image and can be easily confused with bones\cite{nunnari2012hepatic}. Based on these challenges, we aim to retain more abstract and rich features and introduce global feature capture.

\subsection{Overall Architecture }

Given a CT image $\mathbf{X} \in \mathbb{R}^{C \times H \times W}$, where $C$ denotes the number of channels, and with the input being a grayscale image, $C = 1$. $H=512$ and $W=512$ represent the height and width of the image, respectively. Our objective is to segment echinococcosis lesions from the CT image $\mathbf{X}$. Specifically, we propose a novel encoder-decoder model named HES-UNet for segmenting echinococcosis lesions. This model consists of four main components: a multi-scale feature integration encoder (MFSI encoder), a multi-scale global feature filtering module (MGF module), a progressive fusion decoder (PF decoder), and a deep supervision module (DS module), as shown in Fig.~\ref{fig1:env}.

The overall pipeline of HES-UNet can be summarized as follows: first, the MFSI encoder is employed to extract features from the CT image $\mathbf{X}$, generating a set of multi-scale features $\{\mathbf{E}_i \in \mathbb{R}^{(32\times 2^{i-1}) \times \frac{H}{2^i} \times \frac{W}{2^i}} \mid i \in [1,5]\}$, where each scale is processed by an encoder block (EB) as shown in Fig.~\ref{fig2:env}A.
Next, the set $\{\mathbf{E}_i\}$ is fed into the multi-scale aggregation block (MAB) to capture the global feature representation $\mathbf{F}\in \mathbb{R}^{(32\times 2^5) \times \frac{H}{2^5} \times \frac{W}{2^5}}$. 
%Subsequently, $\mathbf{F}$ is input into the multi-scale upsampling block (MUB), 
Subsequently, $\mathbf{F}$ is fed into the MGF module, starting with the multi-scale upsampling block (MUB), resulting in multi-scale global features $\{\mathbf{G}_i \in \mathbb{R}^{(32\times 2^{i}) \times \frac{H}{2^{i+1}} \times \frac{W}{2^{i+1}}} \mid i \in [1,5]\}$, which selectively emphasize different scales of features $\{\mathbf{E_i}\}$. Then, $\mathbf{G_5}$ is passed into the PF decoder to estimate the prediction at the finest scale, $\mathbf{P_5}\in \mathbb{R}^{1 \times \frac{H}{2^5} \times \frac{W}{2^5}}$. Additionally, $\mathbf{E_5}$, $\mathbf{P_5}$, and $\mathbf{G_5}$ are used as inputs to the MGF module, which contains several global attention modules (GAMs) to filter redundant features, thereby generating refined features $\{\mathbf{Q}_i \in \mathbb{R}^{(32\times 2^{i-1}) \times \frac{H}{2^i} \times \frac{W}{2^i}} \mid i \in [1,5]\}$. Finally, $\mathbf{Q_5}$ is combined with the output $\mathbf{D_6}\in \mathbb{R}^{(32\times 2^{4}) \times \frac{H}{2^5} \times \frac{W}{2^5}}$ from the PF decoder and serves as the input to the next-level decoder block, namely "DB 5", to generate the prediction $\mathbf{P_4}\in \mathbb{R}^{1 \times \frac{H}{2^4} \times \frac{W}{2^4}}$. The structure of the decoder block (DB) is shown in Fig.~\ref{fig2:env}B. By iteratively applying steps 5 and 6, we ultimately obtain the final prediction $\mathbf{P_0}\in \mathbb{R}^{1 \times H \times W}$ that matches the original size of the CT image $\mathbf{X}$.

\subsection{Multi-scale Feature Integration Encoder}
To more effectively aggregate multi-scale feature information from input CT image $\mathbf{X}$, we designed an encoder structure named the multi-scale feature integration encoder (MFSI encoder), which consists of 5 encoder blocks (EBs) and a multi-scale aggregation block (MAB). 

Encoder blocks (EBs) capture both local and contextual information while progressively downsampling the feature maps. Each EB consists of convolutional layers, attention layers, and downsampling layers. In the convolutional layers, we use the GHPA~\cite{egeunet} module to extract deep features, achieving multi-axis attention-like effects with linear computational complexity. However, since GHPA relies on depth-wise separable convolution, it is less effective in shallow layers compared to standard 2D convolution. Therefore, we use a 2D convolution with a kernel size of 3 in $\text{EB}_1$ to capture fundamental spatial features, and replace the convolution in $\text{EB}_2$ to $\text{EB}_5$ with the GHPA module. For the attention layers, we introduce the CBAM~\cite{woo2018cbam} to extract channel and spatial attention information. In the downsampling layers, since spatial information is crucial for the HE lesion segmentation task, inspired by HWD~\cite{hwd,li2022efficient}, we designed a multi-directional downsampling block (MDB) capable of preserving rich, multi-directional features during downsampling, as shown in Fig.~\ref{fig2:env}E. Let $\mathbf{M} \in \mathbb{R}^{C \times H \times W}$ represent the input feature map to the MDB. After applying the Haar wavelet transformation, we obtain four feature maps: $\{\mathbf{M}_L, \mathbf{M}_{HV}, \mathbf{M}_{HH}, \mathbf{M}_{HD}\} \in \mathbb{R}^{C \times \frac{H}{2} \times \frac{W}{2}}$, where $\mathbf{M}_L$ retains approximate image information, and $\mathbf{M}_{HV}$, $\mathbf{M}_{HH}$, and $\mathbf{M}_{HD}$ capture vertical, horizontal, and diagonal detail information, respectively. We concatenate these four components to obtain $\mathbf{M}' \in \mathbb{R}^{4C \times \frac{H}{2} \times \frac{W}{2}}$, compressing the spatial dimensions into the channel dimension. Next, we use a $1 \times 1$ convolutional layer to filter spatial information in the channel dimension and match the channel dimension with that of $\mathbf{M}$. Finally, we apply batch normalization and the ReLU activation function to produce the output feature map of MDB, denoted as $\overline{\mathbf{M}} \in \mathbb{R}^{C \times \frac{H}{2} \times \frac{W}{2}}$.

Multi-scale aggregation block (MAB) aggregates multi-scale feature information, as shown in Fig.~\ref{fig2:env}C. To be specific, for the output of each encoder block (EB) $\{\mathbf{E}_i \mid i \in [1,5]\}$, adaptive average pooling is applied to adjust the feature map sizes so that feature maps of different scales match in dimension. Then, a $1 \times 1$ convolution layer is used to filter the features further, denoted as $\{\mathbf{\overline{E}_i}\mid i \in [1,5]\}$ . Additionally, to capture more abstract information, we process $\mathbf{E_5}$ through both the GHPA and CBAM modules, yielding $\mathbf{E_6}$ . Finally, we perform matrix addition to directly combine $\mathbf{E_6}$ with $\{\mathbf{\overline{E}_i}\}$ , resulting in a global feature $\mathbf{F}$ that integrates features from all five encoder levels and a deeply fused global feature representation.

\subsection{Multi-scale Global Feature Filtering Module}
To filter redundant features, we propose the multi-scale global feature filtering module (MGF module), which consists of a module called multi-scale upsampling block (MUB) for learning deep abstract features of the image and 5 global attention modules (GAMs) for aggregating multi-scale image features \cite{li2024iianet}.

To enhance the model’s capability to represent complex image features, we propose a multi-scale upsampling block (MUB), which is a convolutional module incorporating pixel shuffle techniques (Fig.~\ref{fig2:env}G). The specific architecture is shown in Fig.~\ref{fig2:env}D. Specifically, MUB takes the global feature representation $\mathbf{F}\in \mathbb{R}^{1024 \times 16 \times 16}$. as input. First, we apply two consecutive GHPA modules to expand the channel dimension to $4096$. Next, pixel shuffle is used to restore the channels to $1024$ while doubling the width and height, achieving an upsampling effect. Following this, $\mathbf{F}$ passes through a set of convolutional and bilinear interpolation layers to adjust the feature map size for input to the GAM. The output of MUB, denoted as  $\{\mathbf{G}_i \in \mathbb{R}^{(32\times 2^{i}) \times \frac{H}{2^{i+1}} \times \frac{W}{2^{i+1}}} \mid i \in [1,5]\}$ , selectively enhances features of $\mathbf{E_i }$ at various scales. Additionally, to improve gradient flow, we add two extra paths: one using the GHPA without pixel shuffle, and another direct connection path to preserve the initial information in the feature map and prevent information loss. 

To enable efficient multi-scale information interaction, as illustrated in Fig. ~\ref{fig2:env}F, we introduce a series of global attention modules (GAMs). To aggregate features from different levels, we employed 5 GAMs. For each GAM\(_i\) , three inputs are required: low-level features $\mathbf{E}_i$, global features $\mathbf{G}_i$, and intermediate prediction results $\mathbf{P}_i$. First, standard 2D convolution and bilinear interpolation are applied to adjust the size of $\mathbf{G}_i$  to match $\mathbf{E}_i$ . Subsequently, $\mathbf{E}_i$ and $\mathbf{G}_i$  are divided into four groups along the channel dimension, denoted as $\{\mathbf{E}_{ij},\mathbf{G}_{ij}\in \mathbb{R}^{(32\times 2^{i-1}) \times \frac{H}{2^i} \times \frac{W}{2^i}} \mid i \in [1,5], j \in[1,4]\}$. Next, $\mathbf{E}_{ij}$, $\mathbf{G}_{ij}$, and $\mathbf{P}_i$ are concatenated along the channel dimension, forming four sets of mixed features. Each group is subsequently processed through layer normalization and dilated convolution to extract multi-scale features, yielding $\{\mathbf{Q'}_{ij}\in \mathbf{R}^{(32\times 2^{i-2}+1) \times \frac{H}{2^i} \times \frac{W}{2^i}} \mid i \in [1,5], j \in[1,4]\}$. Finally, the feature information is fused through channel concatenation followed by a $1 \times 1$ convolution, producing the output denoted as $\{\mathbf{Q}_i \in \mathbb{R}^{(32\times 2^{i-1}) \times \frac{H}{2^i} \times \frac{W}{2^i}} \mid i \in [1,5]\}$.

\subsection{Progressive Fusion decoder}
The progressive fusion decoder consists of 6 consecutive decoder blocks (DBs), each designed as shown in Fig.~\ref{fig2:env}B. First, DB6 receives $\mathbf{G}_5$ from the MUB. We then apply GHPA and bilinear interpolation to $\mathbf{G}_5$, with the result serving as the output of DB6, denoted as $\mathbf{D}_6$. Simultaneously, $\mathbf{D}_6$ is passed through a $1 \times 1$ convolution to directly generate the most abstract prediction, $\mathbf{P}_5$. Subsequently, $\mathbf{P}_5$ is passed to GAM5 to obtain refined features $\mathbf{Q}_5$ after filtering redundant features. Then, $\mathbf{Q}_5$ and $\mathbf{D}_6$ are added element-wise and input into the next Decoder Block, DB5. This process is repeated, eventually producing the final lesion region prediction, $\mathbf{P}_0$. It is worth noting that, as in the Encoder Block (EB), in DB1, we replace GHPA with a $3 \times 3$ convolution.

\subsection{Loss Function}
To address the issue of gradient vanishing during neural network training, we adopted deep supervision to compute the loss functions at different stages. The binary cross-entropy (BCE) loss function is commonly used for image binary classification,
\begin{equation}
\text{Bce}(\mathbf{y},\hat{\mathbf{y}}) =  -\frac{1}{N} \sum_{i=1}^{N} \left[ \mathbf{y}_i \log(\hat{\mathbf{y}}_i) + (1 - \mathbf{y}_i) \log(1 - \hat{\mathbf{y}}_i) \right],
\label{bce}
\end{equation}
where $N$ represents the total number of pixels, $\mathbf{y}$ is the ground truth, and $\hat{\mathbf{y}}$ is the predicted image. However, since lesion regions are usually small, the Dice loss function is more effective in handling class imbalance,
\begin{equation}
\text{Dice}(\mathbf{y},\hat{\mathbf{y}}) = 1 - \frac{2 \sum_{i=1}^{N} \mathbf{y}_i \cdot \hat{\mathbf{y}}_i}{\sum_{i=1}^{N} \mathbf{y}_i + \sum_{i=1}^{N} \hat{\mathbf{y}}_i}. \label{dice}
\end{equation}

In summary, our loss function can be expressed by \eqref{dice} and \eqref{bce}, 
\begin{equation}
\mathcal{L} = \sum_{i=0}^{5} (\text{Dice}(\mathbf{L}_i,\mathbf{P}_i)+\text{Bce}(\mathbf{L}_i,\mathbf{P}_i))  \times \lambda_i.
\label{L}
\end{equation}
The $\lambda_i$ is set to \{0.1, 0.2, 0.3, 0.4, 0.5, 1\} to balance the losses from different stages of the network.

\begin{figure*}
\centering
\includegraphics[width=0.9\textwidth]{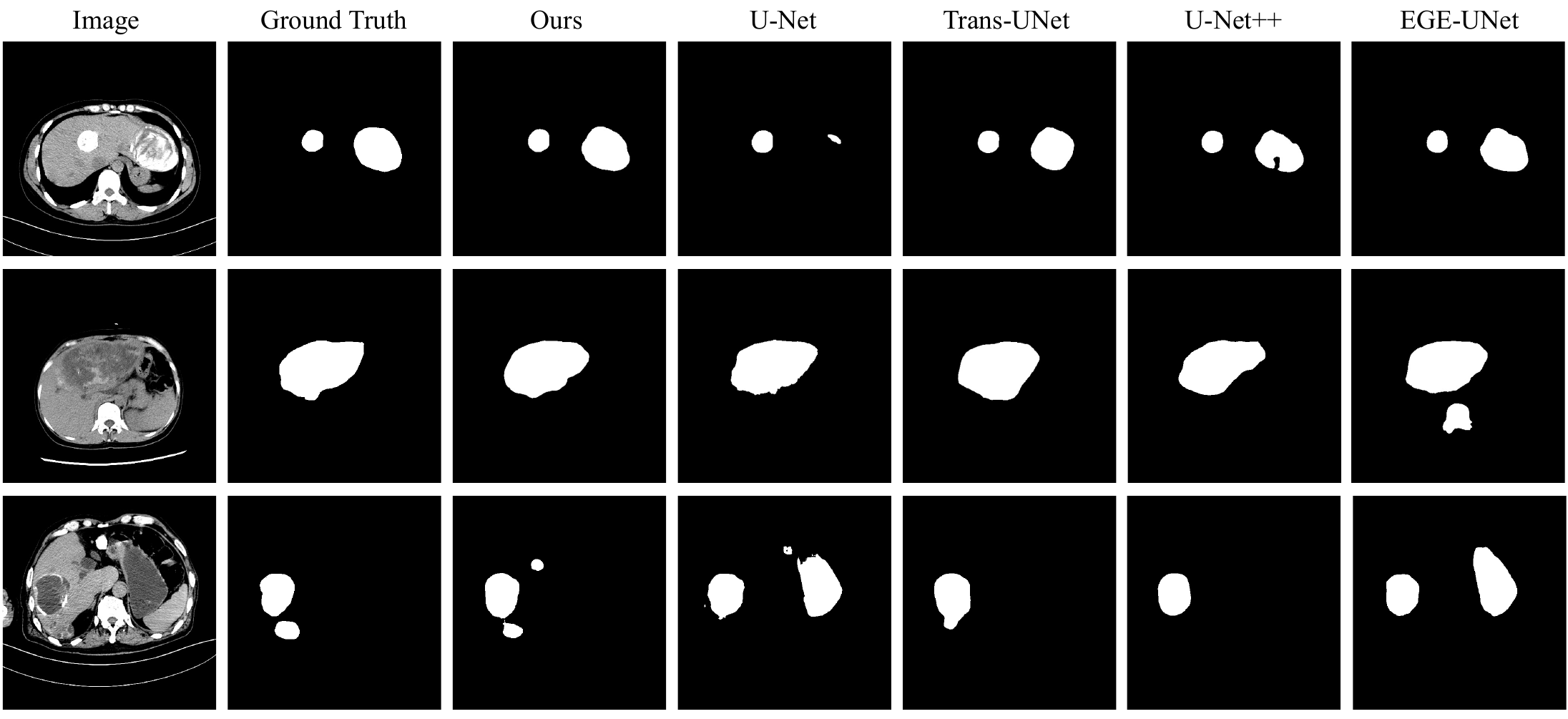}
\caption{Our segmentation results compared with other models.}
\label{fig:result_compar}
\end{figure*}

\section{Experiments and Results}
\subsection{Dataset and Implementation Details}
\textbf{Dataset. }
We collected CT image data from 268 patients with HE (including 137 cases of CE and 131 cases of AE) from a certain hospital and performed preprocessing on the dataset. First, to protect patient privacy, we removed all metadata containing patient identification information. Next, the original grey values (HU values) of the CT images are very broad, exceeding the grey levels that the human eye and display can perceive simultaneously. Therefore, we performed windowing to better highlight the details of different tissues and lesions. Under the guidance of three professional physicians, we selected a window width of 150 HU and a window level of 35 HU to standardize the HU values to the range of [-150, 35]. This value was chosen based on the considerations of liver contours, lesion areas, and bone information. Subsequently, we saved the CT slice data, using 80\% for the training set and 10\% each for the testing and validation sets.

\textbf{Implementation Details. }
As the input consisted of grayscale images, the model's input channel was set to $1$, and the output channel, corresponding to the number of classes, was also set to $1$. The number of channels in the feature maps output by each EB was set to [32, 64, 128, 256, 512], respectively. 

We set the total number of epochs to 200 and employed an early stopping mechanism to monitor the minimum loss on the validation set. If no improvement in validation loss was observed for 50 epochs, the training was halted to prevent overfitting. We selected AdamW\cite{AdamW} as the optimizer, with an initial learning rate of 0.001, and utilized the ReduceLROnPlateau scheduler to reduce the learning rate by half if no improvement in validation loss was seen for 5 consecutive epochs. The batch size was set to 4. Additionally, since obtaining sufficient data from patients with hepatic echinococcosis was challenging, we worked with a relatively small dataset. We applied data augmentation techniques to prevent overfitting, including random image rotations, random scaling, random Gaussian smoothing, and random Gaussian noise.

\subsection{Comparison with State-of-the-art Methods}
We conducted extensive experiments on the dataset to quantitatively compare the segmentation performance of our proposed HES-UNet with some existing medical image segmentation models. The results are shown in Table 1. We selected typical models already showing good segmentation performance, including U-Net \cite{unet}, U-Net++ \cite{unet++}, TransUNet \cite{transunet}, Swin-UNet \cite{swinunet}, Res-UNet \cite{diakogiannis2020resunet}, and EGE-UNet \cite{egeunet}.

HES-UNet achieved improvements of 2.63\%, 1.56\%, and 1.09\% in the DSC metric compared to CNN-based segmentation methods (U-Net and U-Net++), the Transformer-based Swin-UNet, and the hybrid architecture TransUNet, respectively. Additionally, our model demonstrates advantages in precision and recall metrics. These experimental results indicate that the proposed HES-UNet performs excellently in liver lesion segmentation tasks, surpassing other models in segmentation accuracy. Fig.~\ref{fig:result_compar} displays some of the segmentation results.

\subsection{Ablation Analysis}

% In this section, we analyze the MUB, MDB, and MAB modules to demonstrate the impact of different modules on the overall performance. The detailed results of the ablation study are shown in Tab.~\ref{tab:ablation}. The results indicate that incorporating the MUB, MDB, and MAB modules enhances the model’s performance, validating the effectiveness of our proposed model. Compared to the vanilla U-Net, the DSC improved by 3.61\%.

To evaluate the individual contributions of the MUB, MDB, and MAB modules to our model's performance, we conducted an ablation study using the vanilla U-Net as the baseline. The detailed results are presented in Table~\ref{tab:ablation}. The findings demonstrate that integrating the MUB, MDB, and MAB modules significantly enhanced the model's performance, confirming the effectiveness of our proposed architecture. 
Specifically, the DSC, precision, and recall metrics achieved improvements of 1.55\%, 1.79\%, and 1.25\%, respectively, over the baseline.

\begin{table}
\begin{center}
\caption{The comparison results between our HES-UNet and other models. The optimal values are highlighted in bold.} \label{tab:comparison}
\begin{tabular}{c|ccc}
  \toprule
  \textbf{Method} & \textbf{DSC (\%)} & \textbf{Precision (\%)} & \textbf{Recall (\%)} \\
  \midrule
  U-Net\cite{unet}          & 86.58 & 86.05 & 87.12 \\
  U-Net++\cite{unet++}        & 86.76 & 86.73 & 86.34 \\
  TransUNet\cite{transunet}      & 88.12 & 87.76 & 87.98 \\
  Swin-UNet\cite{swinunet}      & 87.65 & 85.62 & 87.24 \\
  Res-UNet\cite{diakogiannis2020resunet}       & 86.87 & 86.05 & 85.43 \\
  EGE-UNet\cite{egeunet}       & 85.81 & 85.29 & 86.34 \\
  \midrule
  HES-UNet \textbf{(Ours)}  & \textbf{89.21} & \textbf{88.14} & \textbf{89.60} \\
  \bottomrule
\end{tabular}
\end{center}
\end{table}

\begin{table}
\begin{center}
\caption{Ablation study of different modules. The optimal values are highlighted in bold.} \label{tab:ablation}
\begin{tabular}{ccc|ccc}
  \toprule
  \textbf{MDB} & \textbf{MUB} & \textbf{MAB} & \textbf{DSC (\%)} & \textbf{Precision (\%)} & \textbf{Recall (\%)} \\
  \midrule
   $\times$ & $\times$ & $\times$ & 86.58 & 86.05 & 87.12\\
  \midrule
  \checkmark & $\times$ & $\times$ & 87.04  & 86.32 & 86.97 \\
  $\times$ & \checkmark & $\times$ & 86.83 & 86.12 & 87.21 \\
  $\times$ & $\times$ & \checkmark & 87.32 & 86.55& 86.73\\
    \midrule
  \checkmark & \checkmark & $\times$ &  87.13& 86.62& 87.74\\
  \checkmark & $\times$ & \checkmark & 87.46& 87.29& 87.35\\
  $\times$ & \checkmark & \checkmark &  87.36& 87.63& 87.92\\
    \midrule
  \checkmark & \checkmark & \checkmark & \textbf{88.13}& \textbf{87.84}& \textbf{88.37}\\
  \bottomrule
\end{tabular}
\end{center}
\end{table}

\section{Conclusion}

In this paper, we proposed an efficient and accurate segmentation model, HES-UNet, for the problem of insufficient multi-scale feature fusion in the segmentation of hepatic echinococcosis (HE) lesions. The model effectively combines convolutional layers and attention mechanisms by introducing an MFSI encoder, MGF module, PF decoder and DS module, which can simultaneously capture local and global features of the image. On the CT dataset of 268 patients collected by us, HES-UNet demonstrated excellent performance, with an overall Dice similarity coefficient (DSC) of 89.21\%. This result is 1.09\% higher than that of TransUNet. Through comparative experiments and ablation analysis with various advanced segmentation models, we verified the effectiveness of the proposed modules and the overall model in improving segmentation accuracy. 
HES-UNet achieves remarkable results in HE lesion segmentation, providing a powerful auxiliary tool for diagnosing HE in resource-poor areas.

\bibliographystyle{IEEEbib}
\bibliography{icme2025references}

\end{document}